\documentstyle[proceedings,numreferences]{crckapb}
\def\be{\begin{equation}}
\def\ee{\end{equation}}
\def\bea{\begin{eqnarray}}
\def\eea{\end{eqnarray}}

\def\sm{M_\odot}
\def\etal{et al. }

\def\be{\begin{equation}}
\def\ee{\end{equation}}
\def\bea{\begin{eqnarray}}
\def\eea{\end{eqnarray}}
\def\sm{M_\odot}
\def\etal{et al. }

\begin{opening}
\title{PRIMORDIAL BLACK HOLES AS A PROBE OF THE EARLY
UNIVERSE AND A VARYING GRAVITATIONAL CONSTANT }

\author{B. J. CARR}
\institute{Astronomy Unit, Queen Mary \& Westfield College, \\
Mile End Road, London E1 4NS, UK}

\end{opening}

\runningtitle{ }

\begin{document}

\begin{abstract}
We discuss recent developments in the study of primordial black holes, focussing
particularly on their formation and quantum evaporation. Such studies can place important constraints on models of the early Universe. An especially interesting
development has been the realization that such constraints may be severely modified if the value of the gravitational
``constant" $G$ varies with cosmological epoch, a possibility which arises in many scenarios
for the early Universe. The nature of the modification depends upon whether the value of $G$ near a black
hole maintains the value it had at its formation epoch (corresponding to gravitational memory) or
whether it tracks the background cosmological value. This is still uncertain but we discuss various
approaches which might help to resolve the issue.
\end{abstract}

\section{Introduction}

It is well known that primordial black holes (PBHs) could have formed in the early Universe \cite{rf:1,rf:2}. A 
comparison of the cosmological density at any time after the Big Bang with the density associated with a black hole
shows that PBHs would have of order the particle horizon mass at their formation epoch:
\be
M(t) \approx {c^3 t\over G} \approx 10^{15}\left({t\over 10^{-
23} \ {\rm s}}\right) g.
\ee
PBHs could thus span an enormous mass range: those formed at the
Planck time ($10^{-43}$s) would have the Planck mass ($10^{-5}$g),
whereas those formed at 1~s would be as large as $10^5\sm$,
comparable to the mass of the holes thought to reside in galactic nuclei. 
PBHs would could arise in various ways \cite{rf:3}. Since the early Universe is unlikely to have been exactly
Friedmann,  they would form
most naturally from initial inhomogeneities but they might also form through other mechanisms at a cosmological phase
transition.
 
The realization that small PBHs might exist prompted Hawking
to study their quantum properties. This led to his famous 
discovery \cite{rf:4}
that black holes radiate thermally with a temperature
\be
T = {\hbar c^3\over 8\pi GMk} \approx 10^{26} \left({M\over g}\right)^{-1}\ {\rm K}
\approx 10^{-7} \left({M\over \sm}\right)^{-1}\ {\rm K}.
   \ee
This means that they evaporate on a timescale 
\be
      \tau(M) \approx {\hbar c^4\over G^2M^2}
\approx 10^{64} \left({M\over \sm}\right )^3~\mbox{y}.
   \ee
Only black holes smaller
than $10^{15}$g would have evaporated by the present epoch, so eqn (1) implies
that this effect could be important only for black holes which formed before $10^{-23}$s. Despite the conceptual importance of this
result, it is bad news for PBH enthusiasts. For since PBHs
with a mass of $10^{15}$g, which evaporate at the present epoch,
would have a temperature of order 100~MeV, the observational 
limit on
the $\gamma$-ray background intensity at 100~MeV immediately 
implies
that their density could not exceed $10^{-8}$ times the
critical density \cite{rf:43}. Not only does this render 
PBHs
unlikely dark matter candidates, it also implies that there is 
little
chance of detecting black hole explosions at the present epoch \cite{rf:90}.

Despite this conclusion, PBH
evaporations could still have interesting cosmological 
consequences. In particular, they might
generate the microwave background \cite{rf:91} or
modify the standard cosmological nucleosynthesis scenario \cite{rf:92} 
or contribute to the cosmic baryon asymmetry \cite{rf:93}.
PBH evaporations might also account for
the annihilation-line radiation coming from the Galactic centre \cite{rf:94} 
or the unexpectedly high
fraction of antiprotons in cosmic rays \cite{rf:95}. PBH explosions occurring
in an interstellar
magnetic field might also lead to radio bursts \cite{rf:96}. 
Even if PBHs had none of
these consequences, studying such effects leads to strong upper limits on how
many of them could ever have formed and thereby constrains models of the early Universe. Indeed PBHs serve as a probe of times much
earlier than that associated with most other ``relicts" of the Big Bang. While photons decoupled at
$10^6$y, neutrinos at 
$1$~s and WIMPs at $10^{-10}$s, PBHs go all the way back to
the Planck time. Therefore even if PBHs never formed, their non-existence
gives interesting information. 

We review the formation mechanisms and evaporation constraints on PBHs 
in Section 2. Much of this material is also contained in my contribution to the 1996 Chalonge School \cite{rf:97}. However, there have been several 
interesting developments since then and these are covered in Section 3. The remaining
sections will examine
how the PBH constraints are modified if the value of the gravitational ``constant" $G$ was different at early times.

As reviewed in Section 4, this idea has a long history and should no longer be regarded as exotic. 
It arises in various scalar-tensor theories of gravity 
and these are a natural setting for many currently popular models of the early Universe.
 Black hole formation and evaporation could be greatly
modified in variable-$G$ cosmologies, since many of their properties (eg. their Hawking temperature) 
depend explicitly on $G$. However, as emphasized by Barrow \cite{rf:5} and discussed in Section 5, the nature of the modification depends upon whether the PBH preserves 
the value of
$G$ at its formation epoch (corresponding to what is termed ``gravitational memory'') or always maintains the changing background
value. There would be interesting modifications to the cosmological
consequences of PBH evaporations in both cases but they would be more dramatic in the first. Barrow \& Carr \cite{rf:6} considered the implications of these two scenarios in detail and this work has been taken further with Goymer
\cite{rf:7}. We
will review the conclusions of these papers in Section 6 and highlight a particularly 
interesting development in Section 7.

\section{PBH formation and constraints on the early Universe}

One of the most important reasons for studying PBHs is that it enables one to 
place limits on the spectrum of density fluctuations in the early Universe. This
is because, if the PBHs form
directly from density perturbations, the fraction of
regions undergoing collapse at any epoch is determined 
by
the root-mean-square amplitude $\epsilon$ of the fluctuations
entering the horizon at that epoch and the equation of state 
$p=\gamma
\rho~(0< \gamma <1)$. One usually expects a radiation equation of
state ($\gamma =1/3$) in the early Universe. In order to 
collapse
against the pressure, an overdense region must be larger than the
Jeans length at maximum expansion and this is just 
$\sqrt{\gamma}$
times the horizon size. On the other hand, it cannot be larger than the horizon size, else it would
form a separate closed universe and not be part of our Universe \cite{rf:8}. 

This has two important implications. Firstly, PBHs forming at time $t$ should
have of order the horizon mass given by eqn (1).
Secondly, for a region destined to collapse to a PBH, one requires the fractional overdensity at the horizon
epoch,
$\delta$, to exceed $\gamma$. Providing the density 
fluctuations have a Gaussian distribution and are spherically
symmetric, one can infer that the fraction of regions of mass 
$M$ which collapse is \cite{rf:9}
   \be
\beta(M) \sim \epsilon(M) \exp\left[-{\gamma^2\over 
2\epsilon(M)^2}\right]
   \ee
where $\epsilon (M)$ is the value of $\epsilon$ when the 
horizon
mass is $M$. The PBHs can have an extended mass spectrum only 
if the
fluctuations are scale-invariant (i.e. with $\epsilon$ 
independent of $M$) but this is expected in many scenarios.

The fluctuations required to make the PBHs may either be primordial or they may arise spontaneously at some epoch. 
One natural source of fluctuations would be inflation \cite{rf:20,rf:21} and, in this
context, $\epsilon(M)$ depends implicitly on the inflationary potential. PBHs
formed before inflation would be drastically diluted but new ones
could form from the fluctuations generated after inflation. Many people have studied PBH formation
in this context \cite{rf:22,rf:23,rf:24,rf:25,rf:26,rf:27,rf:28,rf:100,rf:101} as an important way of constraining the inflationary potential.  
This was the focus of
my 1996 Erice lecture, so it will not be covered again here. Note that the Gaussian assumption has been questioned in the inflationary
context \cite{rf:29,rf:30}, so eqn (4) may not apply, but one still finds that
$\beta$ depends very sensitively on $\epsilon$.

In some situations
eqn (4) would fail qualitatively. For example,  PBHs would form more easily if the
equation of state of the Universe were ever soft ($\gamma\ll$1). This might apply if there was a phase 
transition which channelled the mass of the Universe into non-relativistic 
particles or which temporally reduced the pressure. In this case, only those regions which are sufficiently spherically
symmetric at maximum expansion can undergo collapse; the dependence of $\beta$ on $\epsilon$ would then have the form \cite{rf:19}
\be
\beta = 0.02\epsilon^{13/2},
\ee
which is much weaker than indicated by eqn (4), but there would still be a
unique relationship between the two parameters.
Some formation mechanisms for
PBHs do not depend on having primordial
fluctuations at all.  For example,  at any spontaneously broken symmetry epoch, PBHs might form 
through the collisions of bubbles of broken symmetry \cite{rf:31,rf:32,rf:33}. 
PBHs might also form spontaneously through the
collapse of cosmic strings \cite{rf:34,rf:35,rf:36,rf:37,rf:38} or domain walls \cite{rf:99}. In these
cases $\beta(M)$  depends, not on $\epsilon(M)$, but
on other cosmological parameters, such the bubble formation
rate or the string mass-per-length. These mechanisms were discussed in 
more detail in my 1996 Erice contribution \cite{rf:97}.

In all these scenarios, the current density parameter $\Omega_{PBH}$ associated
with PBHs which form at a redshift $z$ or time $t$ is related to
$\beta$ by \cite{rf:9}
   \be
\Omega_{\rm PBH} = \beta\Omega_R(1+z) \approx 10^6 
\beta\left({t\over s}\right)^{-1/2} \approx 10^{18}\beta\left({M\over 10^{15}g}\right)^{-1/2}
   \ee
where $\Omega_R \approx 10^{-4}$ is the density parameter of the microwave
background and we have used eqn (1). The $(1+z)$ factor arises 
because the radiation density scales as $(1+z)^4$, whereas the PBH
density scales as $(1+z)^3$. Any limit on $\Omega_{PBH}$ therefore places a constraint on $\beta(M)$ and 
the constraints are summarized in Fig. 1.
The constraint for non-evaporating mass ranges above
$10^{15}$g comes from requiring
$\Omega_{PBH}<1$.
Stronger constraints are associated with PBHs
smaller than this since they would have evaporated by now \cite{rf:42,rf:44,rf:45,rf:46}. 
The strongest one is the $\gamma$-ray limit associated with the $10^{15}$g PBHs evaporating
at the present epoch \cite{rf:43}. Other ones are associated with the generation of entropy and modifications to the cosmological production of light 
elements. The
constraints below $10^{6}$g are based on the (not necessarily secure) assumption that evaporating PBHs leave stable Planck
mass relics, in which case these  relics are required to have less than the critical density \cite{rf:23,rf:39,rf:40,rf:102}.

The constraints on 
$\beta(M)$ can be converted into constraints on 
$\epsilon(M)$ using eqn (4) and these are shown in Fig. 2. Also shown here are the (non-PBH) constraints
associated with the spectral distortions in the cosmic microwave background induced by the dissipation of 
intermediate scale density perturbations and the COBE quadrupole measurement, as well as lines corresponding
to various slopes in the $\epsilon (M)$ relationship. This shows that  one needs the
fluctuation amplitude to decrease with increasing scale in order to produce PBHs. 

\begin{figure}\label{F1}
\vspace{3.3in}
\caption{Constraints on $\beta(M)$}
\end{figure}

\begin{figure}\label{F2}
\vspace{3.3in}
\caption{Constraints on $\epsilon(M)$}
\end{figure}

\section{Recent Developments}

Recent hydrodynamical calculations for the $\gamma=1/3$ case have refined the criterion $\delta > \gamma$ for PBH formation and this modifies the estimate for $\beta(M)$ given by eqn (4).
Niemeyer \& Jedamzik \cite{rf:10} find that one needs $\delta >0.8$ rather than $\delta >0.3$ to ensure PBH 
formation, and Shibata \& Sasaki \cite{rf:11} reach similar conclusions. They also find that there is little accretion after PBH formation, as expected theoretically  \cite{rf:8}.

Another interesting development has been the application of ``critical phenomena"
to PBH formation.  Studies of the collapse of various types of spherically symmetric matter fields have shown that there
is  always a critical solution which separates those configurations which form a black hole from 
those which disperse to an asymptotically flat state. The configurations are described by some index $p$ and, as the
critical index $p_c$ is approached, the black hole mass is found to scale as $(p-p_c)^{\eta}$ for some exponent
$\eta$. This effect was first discovered for scalar fields \cite{rf:12} but subsequently
demonstrated for radiation \cite{rf:13} and then more general fluids with equation of state $p=\gamma \rho$
\cite{rf:14,rf:15}. 

In all these studies the spacetime was assumed to be asymptotically flat. However, Niemeyer \& Jedamzik \cite{rf:16} have 
recently applied
the same idea to study black hole formation in asymptotically Friedmann models and have found similar results. 
For a variety of initial density perturbation profiles, they find that the
relationship between the PBH mass and the the horizon-scale density perturbation has the form
\be
M = K M_H(\delta - \delta_c)^{\gamma}
\ee
where $M_H$ is the horizon mass and the constants are in the range $0.34<\gamma<0.37$, $2.4<K<11.9$ and
$0.67<\delta_c <0.71$ for the various configurations. Since $M \rightarrow 0$ as $\delta \rightarrow \delta_c$, 
this
suggests that PBHs may be much smaller than the particle horizon at formation (although it is clear that a fluid
description must break down if they are too small) and it also modifies the mass spectrum \cite{rf:17,rf:18}.

There has been particular interest recently in whether PBHs could have 
formed at the quark-hadron phase transition at $10^{-5}$s. This is because the horizon mass is of order $1\sm$ then, so
such PBHs would naturally have the sort of mass required to explain the MACHO microlensing results \cite{rf:51}.  
This is discussed in more detail in my other lecture at this meeting. One might expect PBHs to form more easily at that epoch because of a temporary softening of the equation of 
state. If the QCD phase 
transition is assumed to be of 1st order, then hydrodynamical calculations show that the value of
$\delta$ required for PBH formation is indeed reduced below the value which pertains in the radiation case \cite{rf:52}. 
This means that 
PBH formation will be strongly enhanced at the QCD epoch, with the mass distribution being peaked around the
horizon mass. 

One of the interesting
implications of the PBH MACHO scenario is the possible existence of a halo population of {\it binary} black
holes \cite{rf:53}. With a full halo of such objects, there could then be $10^8$ binaries inside 50 kpc and some of these could be coalescing due to
gravitational radiation losses at the present epoch \cite{rf:54}. Current interferometers
(such as LIGO) could detect such coalescences within 50 Mpc, corresponding to a few events per year. Future
space-borne interferometers (such as LISA) might detect 100 coalescences per year. If the associated gravitational
waves were detected, it would provide a unique probe of the halo distribution (eg. its density
profile and core radius \cite{rf:55}.

Kohri \& Yokoyama
\cite{rf:41} have recently improved the constraints on $\beta(10^8 - 10^{10}g)$ which come from cosmological
nucleosynthesis considerations. Constraints from neutrino background
have also been presented by Bugaev \& Konischev \cite{rf:98}. The recent detection of a Galactic $\gamma$-ray background
\cite{rf:47},  measurements of the antiproton flux \cite{rf:48}, and the discovery of very short period
$\gamma$-ray  burts \cite{rf:49} may even provide positive evidence for
such PBHs. This is discussed in detail elsewhere \cite{rf:50}.

Some people have emphasized the possibility of detecting
very high energy cosmic rays from PBHs using air shower techniques \cite{rf:103,rf:104}. However, recently these 
efforts have been set back by the claim of Heckler \cite{rf:105} that QED interactions could
produce an optically thick photosphere once the black hole temperature exceeds $T_{crit}=45$ GeV. In this case, the mean photon
energy is reduced to $m_e(T_{BH}/T_{crit})^{1/2}$, which is well below $T_{BH}$, so
the number of high energy photons is much reduced. He
has proposed that a similar effect may operate at even lower temperatures due to QCD effects \cite{rf:106}. This is discussed further in the contribution of Kapusta 
at this meeting \cite{rf:107}. However, these 
arguments should not be regarded as definitive: MacGibbon \etal \cite{rf:108} claim that Heckler has not included Lorentz factors correctly
in going from the black hole frame to the centre-of-mass frame of the interacting particles; in their calculation QED interactions
are never important. 

\section{Cosmology in varying-G theories}

Most variable-G scenarios associate the gravitational ``constant'' with some form of scalar field
$\phi$. This notion has its roots in Kaluza-Klein theory, in which a scalar field appears
in the metric component $g_{55}$ associated with the 5th dimension. 
Einstein-Maxwell theory then requires that this field be related to $G$ \cite{rf:56}. Although this was assumed constant
in the original Kaluza-Klein theory,
Dirac \cite{rf:57} noted the the ratio of the electric to gravitational force between to protons ($e^2/Gm_p^2$)
and the ratio of the age of the Universe to the atomic timescale ($t/t_a$) and the square-root of the number of
particles in the Universe ($\sqrt{M/m_p}$) are all comparable and of order $10^{40}$. This unlikely coincidence led him
to propose that these relationships must {\it always} apply, which requires 
\be
G \propto t^{-1}, \;\;\;GM/R \sim 1,
\ee
where $R\sim ct$ is the horizon scale. The first condition led Jordan \cite{rf:58} to propose a theory in which the scalar
field in Kaluza-Klein theory is a  function of {\it both} space and time, and this then implies that $G\sim \phi^{-1}$ has
the same property. The second condition
implies the Mach-type relationship
$\phi \sim M/R$, which suggests \cite{rf:59} that $\phi$ is a solution of the wave equation $\Box \phi \sim \rho$. This
motivated Brans-Dicke (BD) theory \cite{rf:60}, in which the Einstein-Hilbert Lagrangian is replaced by
\be
L = \phi R - \frac{\omega}{\phi}\phi_{,\mu}\phi_{,\nu}g^{\mu \nu} + L_m \;,
\ee
where $L_m$ is the matter Langrangian and the constant $\omega$ is the BD parameter. The potential $\phi$
then satisfies 
\be
\Box \phi = \left(\frac{8\pi}{2\omega +3}\right) T,
\ee
where $T$ is the trace of the matter stress-energy tensor, and this has the required Machian form. Since $\phi$ must
have a contribution from {\it local} sources of the form
$\Sigma_i (m_i/r_i)$, this entails a violation of the Strong Equivalence Principle. In order to test this, the PPN
formalism was introduced. Applications of this test in a variety of astrophysical situations (involving the
solar system, the binary pulsar and white dwarf cooling) currently require
$|\omega| > 500$, which implies that the deviations from general relativity can only ever be small in BD
theory \cite{rf:61}.

The introduction of generalized scalar-tensor theories \cite{rf:62,rf:63,rf:64}, in which $\omega$ is itself a function
of $\phi$, led to a considerably broader range of variable-$G$ theories. In  particular, it permitted the possibility that
$\omega$ may have been small at early times (allowing noticeable variations of $G$ then) even if it is large
today. In the last decade interest in such theories has been revitalized as a result of early Universe studies.
Inflation theory \cite{rf:65} has made the introduction of scalar fields almost mandatory and
extended inflation specifically requires a model in which $G$ varies \cite{rf:33}.
In higher dimensional Kaluza-Klein-type cosmologies, the variation in the sizes of the extra dimensions also naturally
leads to a variation in $G$ \cite{rf:66,rf:67,rf:68}. The currently popular
low energy string cosmologies necessarily involve a scalar (dilaton) field \cite{rf:69} and bosonic superstring
theory, in particular, leads \cite{rf:70} to a Lagrangian of the form (9) with $\omega =-1$.

The intimate connection between dilatons, inflatons and scalar-tensor theory arises because one can always transform
from the (physical) Jordan frame to the Einstein frame, in which the Lagrangian has the standard Einstein-Hilbert form
\cite{rf:71}
\be
L = \bar{R} - 2\psi_{,\mu}\psi_{,\nu}\bar{g}^{\mu \nu} + L_m.
\ee
Here the new scalar field $\psi$ is defined by 
\be
\rm{d}\psi = \left(\frac{2\omega+3}{2}\right)^{1/2}\frac{\rm{d}\phi}{\phi}
\ee
and the barred (Einstein) metric and gravitational constant are related to the
original (Jordan) ones by 
\be
g_{\mu \nu} = A(\phi)^2 \bar{g}_{\mu \nu},\;\;\;G = [1+\alpha^2(\phi)] A(\phi)^2 \bar{G}, \quad \alpha \equiv A'/A,
\ee
where the function $A(\phi)$ specifies a conformal transformation and a prime denotes $d/d\phi$. Thus scalar-tensor theory can be related to
general relativity plus a scalar field, although the theories are not identical because
particles do not follow geodesics in the Einstein frame.

The behaviour of homogeneous cosmological models in BD theory is well understood \cite{rf:72}. Their crucial feature is
that they are vacuum-dominated at early times but always tend towards the general relativistic solution during the
radiation-dominated era. This is a consequence of the fact that the radiation energy-momentum tensor is trace-free
[i.e. $T=0$ in eqn (10)]. This
means that the full radiation solution can be approximated by joining a
BD vacuum solution to a general relativistic radiation solution at some time $t_1$, which 
may be regarded as a free parameter of the theory. However, when the matter density becomes greater than the radiation
density at $t_e \sim 10^{11}$s,  
the equation of state becomes that of dust $(p = 0)$ and $G$ begins to vary again. For a $k = 0$ model, one can show 
that in the three eras \cite{rf:6}
\be
G = G_0 (t_0/t_e)^n, \quad a \propto t^{(2-n)/3} \quad (t>t_e)
\ee
\be
G = G_e \equiv G_0 (t_0/t_e)^n , \quad  a \propto t^{1/2} \quad (t_1<t<t_e)
\ee
\be
G = G_e (t/t_1)^{-(n + \sqrt{4n + n^2})/2}, \quad a \propto t^{(2 - n - \sqrt{4n + n^2})/6} \quad (t<t_1)
\ee
where $G_0$ is the value of $G$ at the current time $t_0$, $n \equiv 2/(4 + 3\omega)$ and $(t_0 / t_e) \approx 10^6$. 
Since the BD coupling constant is constrained by $|\omega| > 500$, which implies 
$|n| < 0.001$, eqns (14) to (16) imply that the deviations from general relativity are never large if the value of
$n$ is always the same. However, as we now explain, it is also interesting to consider BD models
in which $n$ and $\omega$ can vary and thus violate the current constraints.

The behaviour of cosmological models in more general scalar-tensor theories depends on the form of $\omega(\phi)$
but they still retain the feature that the general relativistic solution is a late-time attractor during the radiation
era.  Since one requires $G\approx G_0$ to $10\%$ at the epoch of primordial nucleosynthesis \cite{rf:72}, one needs the
vacuum-dominated phase to end at some time $t_v <$ 1 s. The theory approaches general relativity in the weak field limit
only if
$\omega \rightarrow \infty$ and $\omega '/\omega ^3 \rightarrow 0$ but $\omega(\phi)$ is
otherwise unconstrained. Barrow \& Carr consider a toy model in which
\be
2\omega +3 = 2\beta (1-\phi/\phi_c)^{-\alpha}
\ee
where $\alpha$ and $\beta$ are constants. This leads to
\be
2\omega +3 \propto t^{-\alpha/(2-\alpha)}, \quad \omega '/\omega ^3 \propto t^{(1-2\alpha)/(1-\alpha)} \quad (t<t_v),
\ee
so one requires $1/2<\alpha<2$ in order to have $\omega \rightarrow \infty$ and $\omega '/\omega ^3
\rightarrow 0$ as $t \rightarrow \infty$. In the $\alpha =1$ case, one finds
\be
G \propto t^{-2\lambda/(3-\lambda)}, \quad a \propto t^{(1-\lambda)/(3-\lambda)}, \quad \lambda \equiv
\sqrt{3/(2\beta)} \quad \;\;\; (t<t_v).
\ee
During the vacuum-dominated era, such models can therefore be regarded 
as BD
solutions in which $\omega$ is determined by the parameter $\beta$ and unconstrained by any limits on $\omega$ at the
present epoch. After $t_v$, $G$ is constant and one has the standard radiation-dominated or dust-dominated  
behaviour. 

The consequences of the cosmological variation of $G$ for PBH evaporation depend upon how the value of $G$
near the black hole evolves. Barrow \cite{rf:5} introduces two possibilities: in scenario A, $G$ everywhere maintains
the background cosmological value (so $\phi$ is homogeneous); in scenario B, it
preserves the value it had at the formation epoch near the black hole even though it 
evolves at large distances (so $\phi$
becomes inhomogeneous).
On the assumption that a PBH of mass $M$ has a temperature and mass-loss rate
\be
T = (8\pi GM)^{-1},\quad \dot{M} \approx - (GM)^{-2},
\ee
with $G=G(t)$ in scenario A and $G=G(M)$ in scenario B, Barrow \& Carr calculate the evaporation 
time $\tau$ for various values of the parameters $n$ and $t_1$ in BD theory \cite{rf:6}. The results are shown in Fig. 3(a) for
scenario A and Fig. 3(b) for scenario B. Here $M_*$ is the mass of a PBH evaporating at the present
epoch, $M_e$ is the mass of a PBH evaporating at time $t_e$ and $M_{crit}$ is the mass of a PBH
evaporating at the present epoch in the standard (constant $G$) scenario. In scenario A with $n<-1/2$, there is
a maximum mass of a PBH which can ever evaporate and this is denoted by $M_{\infty}$. The results for the
scalar-tensor with
$\omega(\phi)$ given by eqn (17) with
$\alpha=1$ are  shown in Fig. 3(c) for scenario B with various values of the parameters $\lambda$ and $t_v$.  
The corresponding modifications to the constraints on $\beta(M)$ in all three cases are shown in Fig. 3(d), which should be
compared to Fig. 1. 

\begin{figure}\label{F3}
\vspace{4.3in}
\caption{Dependence of the PBH evaporation time $\tau$ on initial mass $M$ in (a) BD theory with scenario A, (b) 
BD theory with scenario B, (c) scalar-tensor theory with $\alpha=1$ and scenario B. Also shown are (d) the modifications to
the constraints on $\beta(M)$ in these cases.}
\end{figure}

\section{Black holes in scalar-tensor theory} 

Barrow \& Carr considered both scenarios A and B but did not attempt
decide which was more plausible. In this section we address this question more
carefully. The main argument for scenario A comes from an important result of 
Hawking \cite{rf:73}. He showed that in BD theory, providing the weak energy condition holds, the gradient 
of $\phi$ must be zero everywhere for stationary, asymptotically flat black holes. This means that such black 
holes are identical to those in general relativity.  This result can be generalized
to all scalar-tensor theories and suggests that such theories are in agreement with the ``no-hair'' theorem.

Numerical calculations support this theorem \cite{rf:74,rf:75,rf:82}. Collapse is
accompanied by outgoing scalar gravitational radiation, which 
radiates away the scalar mass until the black hole settles down to the Schwarzschild form with a constant scalar field. In particular, Harada et al. \cite{rf:76} have investigated Oppenheimer-Snyder collapse in which a ball of dust described by a $k=+1$ Friedmann interior and a Schwarzschild exterior 
collapses to a black hole. In these calculations the scalar field is taken to be constant before the 
collapse and its back-reaction on the metric is assumed to be always negligible. It is found that, as 
the collapse proceeds, a scalar gravitational wave propagates outwards before the scalar 
field settles down to being constant again.

It should be stressed that
the scalar
no hair theorem has only been proved for asymptotically flat spacetimes, so it is not clear that it also applies in the
asymptotically Friedmann case. While the no hair theorem suggests that $\phi$ should tend to a {\it locally} 
constant value (close to the black hole), it is not obvious that this needs to be the asymptotic cosmological value.
Indeed, since the homogeneizing of $\phi$ is only ensured by scalar wave emission, one might infer that this can only be
achieved on scales less than the particle horizon. 

One way to determine what happens is to seek a precise mathematical model
for a  black hole in a cosmological background.  
For example, one can try to match a black hole and cosmological solution over some boundary
$\Sigma$. Such a matching is provided in general  relativity by the Einstein-Straus or ``Swiss cheese"
model \cite{rf:80}. Here a Friedmann exterior is matched with a general spherically symmetric interior.
If there is no scalar field, it turns out that the latter has to be the static Schwarzschild solution but the situation may
be  more complicated in the present context due to the presence of scalar gravitational radiation. 
In general one can show that the following continuity conditions must apply at $\Sigma$:
\be
[g_{\mu \nu}] = 0, \quad [G_{\mu \nu} n^{\mu} n^{\nu}] = [G_{\mu \nu} u^{\mu} n^{\nu}] = 0, 
\quad [\phi] = 0, \quad [\phi_{\mu}n^{\mu} ] = 0
\ee
where $n^{\mu}$ and $u^{\mu}$ are 4-vectors normal and tangent to $\Sigma$, respectively \cite{rf:81}.
Unfortunately, it turns out that an Einstein-Straus type 
solution does not exist in BD theory. This is because the only way to satisfy the 
junction conditions (21) is if $\phi$ is spatially and temporally constant, which is just 
the general relativistic case.

Jacobsen \cite{rf:83} has addressed the problem analytically by looking for a spherically symmetric solution which represents a perturbation of the Schwarzschild solution near
the origin but is asymptotically Friedmann at large distances, with $\phi$ satisfying
the appropriate cosmological conditions. 
He presupposes that the black hole event horizon is much smaller than the particle
horizon, so that the cosmological timescale is much longer than the black hole timescale. In this case, he finds that there is little lag between the
value of $\phi$ at the event horizon and particle horizon, which suggests that  memory can only be weak. However, it must be emphasized 
that this conclusion need not follow if the
black hole has a size {\it comparable} to the particle horizon at formation and, as indicated in Section 2, this is expected for a PBH.

Another way to investigate the problem is to study the collapse of dust in a 
Tolman-Bondi
background using the same approximation employed by Harada et al. \cite{rf:76}, i.e. neglecting the back reaction of the scalar field, but requiring that $\phi$ have the required cosmological time-dependence at large distances.  One puts in an initial density perturbation for the 
dust but assumes that $\phi$
is initially homogeneous.  Our preliminary numerical calculations \cite{rf:109} use the characteristic method to determine the evolution of the scalar field perturbation 
along null and constant-time hypersurfaces. We find that $\phi$ does initially
build up near the centre but it then gets smoothed out, tending eventually to homogeneity. Although this suggests that
there is no gravitational memory, it should be stressed that this 
conclusion only applies for dust and if one neglects the back reaction.

\section{Variations of gravitational memory}

Since we lack definite knowledge about the evolution of the scalar field when a black
hole forms in a cosmological background, it is useful to consider a range of
scenarios which go beyond the two possibilities envisaged by Barrow. In general, the background scalar field will have a present value $\phi(t_0)$ and a value $\phi(t_f)$ when the black hole first formed. However, it is likely to develop inhomogeneties for
at least some intervening period. 
In the following discussion, we will characterize the degree of gravitational memory 
by comparing the value of the scalar field at the black hole event horizon ($\phi_{EH}$) and the cosmological 
particle horizon ($\phi_{PH}$). We first
consider the two extreme situations described by Barrow \cite{rf:5}:

\be
Scenario\;A: \quad \quad \phi_{EH}(t) = \phi_{PH}(t) \quad \rm{for\ all}\ t
\ee
A Schwarzschild black hole forms at time $t_f$ with its event horizon radius being 
$R_f = 2G(t_f)M$. If $G(t)$ evolves with time, then the black hole adjusts quasi-statically 
through a sequence of Schwarzschild states approximated by $R = 2G(t)M$, see Fig. 4(a). 
In this scenario there is no gravitational memory.

\be
Scenario\; B: \quad \quad  \phi_{EH}(t) = \phi_{EH}(t_f) \quad \rm{for\ all}\ t
\ee
A Schwarzschild black hole of size $R_f$ forms at time $t_f$ and, while $G(t)$ equals
the evolving background value beyond some scale-length 
$R_m \geq R_f$, it remains constant within $R_m$, see Fig. 4(b). In this case the black hole size
is determined by $G(t_f)$ even at the present epoch and this means that the 
region $R < R_m$ has a memory of the gravitational ``constant'' at the 
time of its formation.
\vskip .2in

Neither of these scenarios can be completely realistic since they both assume that $\phi$ is homogeneous almost everywhere.  However, even if $\phi$ were homogeneous initially,  
one would expect it to become inhomogeneous as collapse proceeds. Indeed, in the dust case, this is 
confirmed by the numerical calculations described above \cite{rf:109}.  Therefore, if the background value is 
increasing (as usually applies), one would expect
$\phi$ in the collapsing region to become first {\it larger} than the background value on a local dynamical 
timescale and then {\it smaller} than it on a cosmological timescale. Such behaviour would
necessarily entail a variation of $\phi$ in space as well as time. 
We must also
allow for the possibility that
$\phi$ may vary interior to $R_m$ but on a slower or faster timescale than the background. We therefore propose two
further scenarios:
\vskip .2in

\be
Scenario\;C: \quad \quad  |\dot{\phi}_{EH}(t)| \geq |\dot{\phi}_{PH}(t)| \quad \rm{for\ all}\ t
\ee
where the dot represents a time derivative. This 
implies that the scalar field evolves faster at the event horizon than at the 
particle horizon until it eventually becomes homogeneous, see Fig. 4(c). We describe this as
{\it short-term} gravitational memory and it reduces to scenario A as the timescale to become homogeneous tends to zero.
This would apply, for example,
if $\phi$ were to change on the dynamical timescales of the black hole since this is usually less than
the cosmological timescale.

\be
Scenario \;D: \quad \quad  \label{beta}
|\dot{\phi}_{EH}(t)| < |\dot{\phi}_{PH}(t)| \quad \rm{for\ all}\ t
\ee
This implies that $\phi$ evolves faster at the particle horizon than 
the event horizon, see Fig. 4(d). We describe this is as {\it weak} gravitational memory 
and it reduces to scenario B 
when the left-hand-side of eqn (\ref{beta}) is zero. In this case, the evolution of $\phi$ is  
again dominated by the black hole inside some length-scale $R_m$. 
Note that, in either this scenario or the last one, the length-scale $R_m$ need not be fixed, since it could either grow or
shrink as  scalar gravitational radiation propagates. A particular example of this, to which we return shortly, would be 
{\it self-similar} gravitational memory, in which the ratio of $\phi_{EH}$ to $\phi_{PH}$ always remains the same.  
\vskip .2in

\begin{figure}\label{F4}
\vspace{4.0in}
\caption{Different possiible forms for the evolution of the scalar field profile $\phi(r)$ for
(a) no memory, (b) strong memory, (c) short-term memory, (d) weak memory.}
\end{figure}

\section{Gravitational memory and the accretion of a stiff fluid}

In general relativity there is an equivalence between a scalar field and a stiff 
fluid and this can be exploited in studying gravitational memory. In the Einstein frame, the energy
momentum tensor for a perfect fluid is
\be
\bar{T}_{\mu \nu} = (\rho + p)u_{\mu} u_{\nu} + \bar{g}_{{\mu \nu}} p
\ee
where $u_{\mu}$ is the velocity of the 
fluid. If we define a velocity field by
\be
\label{FVel}
u_{\mu} = \frac{\bar{\phi}_{\mu}}{(-\bar{g}^{\rho \sigma} \bar{\phi}_{\rho} \bar{\phi}_{\sigma})^{1/2}}\;,
\ee
this gives
\be
\bar{T}_{\mu \nu} = -\frac{(\rho + p) \bar{\phi}_{\mu} \bar{\phi}_{\nu}}{\bar{g}^{\rho \sigma} \bar{\phi}_{\rho}
\bar{\phi}_{\sigma}} + p \bar{g}_{\mu \nu}\;.
\ee
By comparing this to the energy-momentum tensor for a scalar field, we find that
\be
p = \rho = -\frac{1}{2} \bar{g}^{\rho \sigma} \bar{\phi}_{\rho} \bar{\phi}_{\sigma},
\ee
so we have a stiff fluid. This 
equivalence applies provided that the derivative of the scalar field is timelike. Otherwise the 
velocity field defined in (\ref{FVel}) would be imaginary. 

This is relevant to the gravitational memory problem because we can now interpret the various scenarios 
discussed in Section 6 in terms of the {\it accretion} of a stiff fluid. 
If the
black hole does not accrete at all or accretes very little, this will correspond to 
strong or weak gravitational memory (scenarios B and D, respectively). However, if enough accretion occurs to homogenize
$\phi$, this will correspond to short-term gravitational memory (scenario C). The faster the accretion, the shorter the
memory, so scenario A corresponds to the idealization in which homogenization is instantaneous.

A simple Newtonian treatment \cite{rf:1} for a general fluid suggests that the accretion rate in the Einstein 
frame should be
\be
\dot{M} = 4\pi \rho R^2_A v_s,
\ee
where $R_A= G M/ v_s^2$ is the accretion radius and $v_s$ is the sound-speed in the accreted fluid. For a stiff fluid,
$v_s=c$ and $R_A = G M/c^2$, while $\rho \sim 1/(Gt^2)$ in a Friedmann universe at early times, so we have 
\be
\frac{\rm{d}M}{\rm{d}t} \approx \frac{G M^2}{c^3 t^2}.
\ee
This can be integrated to give
\be
\label{mass}
M \approx \frac{c^3t/G}{1 + \frac{t}{t_f} \left(\frac{c^3t_f}{GM_f} - 1 \right)}
\ee
where $M_f$ is the black hole mass at the time $t_f$ when it formed. If we define a parameter $\eta = GM_f / c^3t_f$,
then eqn (\ref{mass}) implies
\be
M\rightarrow M_f(1 - \eta)^{-1} \quad \rm{as} \quad t\rightarrow \infty.
\ee
If $\eta \ll 1$, the black hole could not grow very much. However, if $\eta$ is close to 
1, which must be the case if $v_s \approx c$, then the black hole could grow significantly. 
In particular, in the limit $\eta =1$, eqn (\ref{mass}) implies $M \sim t$, so the black hole grows 
at the same rate as the universe. This simple calculation suggests that a black hole surrounded by a 
stiff fluid can accrete enough to grow at the same rate as the Universe.

Since the above calculation neglects the effects of the cosmological expansion, one needs a relativistic
calculation to check this. The Newtonian result suggests that one should look for a spherically symmetric
{\it self-similar} solution, in which every dimensionless variable is a function of $z=r/t$, so that it is
unchanged by the transformation 
$t \rightarrow at,\ r \rightarrow ar$ for any constant $a$. This problem has an interesting but rather
convolved history. By looking for a black hole solution attached to an exact Friedmann solution via a sonic point, Carr \&
Hawking first showed that there is no such solution for a radiation fluid \cite{rf:8} and the argument can be 
extended to a general $p=\gamma \rho$ fluid with
$0<\gamma <1$.  Lin et al. \cite{rf:88} subsequently claimed that there is such a solution in the special case $\gamma =1$.
However, Bicknell
\& Henriksen \cite{rf:89} then showed that this solution is unphysical, in that the density gradient diverges at the 
event horizon. This
suggests that the black hole must soon become much smaller than the particle horizon, after which 
eqn (32) implies there will be very little further accretion. Therefore the stiff fluid analysis suggests that there
should be at least {\it weak} gravitational memory.

\section{Conclusions}

We have seen that studying the formation and evaporation of PBHs can
place interesting constraints on models of the early universe even if they never existed. 
On the other hand, if they did exist, PBHs can provide unique information about times much earlier than those probed by any other relics of the Big Bang. In particular, they may provide information about the variation of $G$ at early times. The precise signature of such a variation depends upon the degree to which a black hole can ``remember" the value of $G$ at its formation epoch. This is still unclear but various methods are being pursued to resolve this issue.

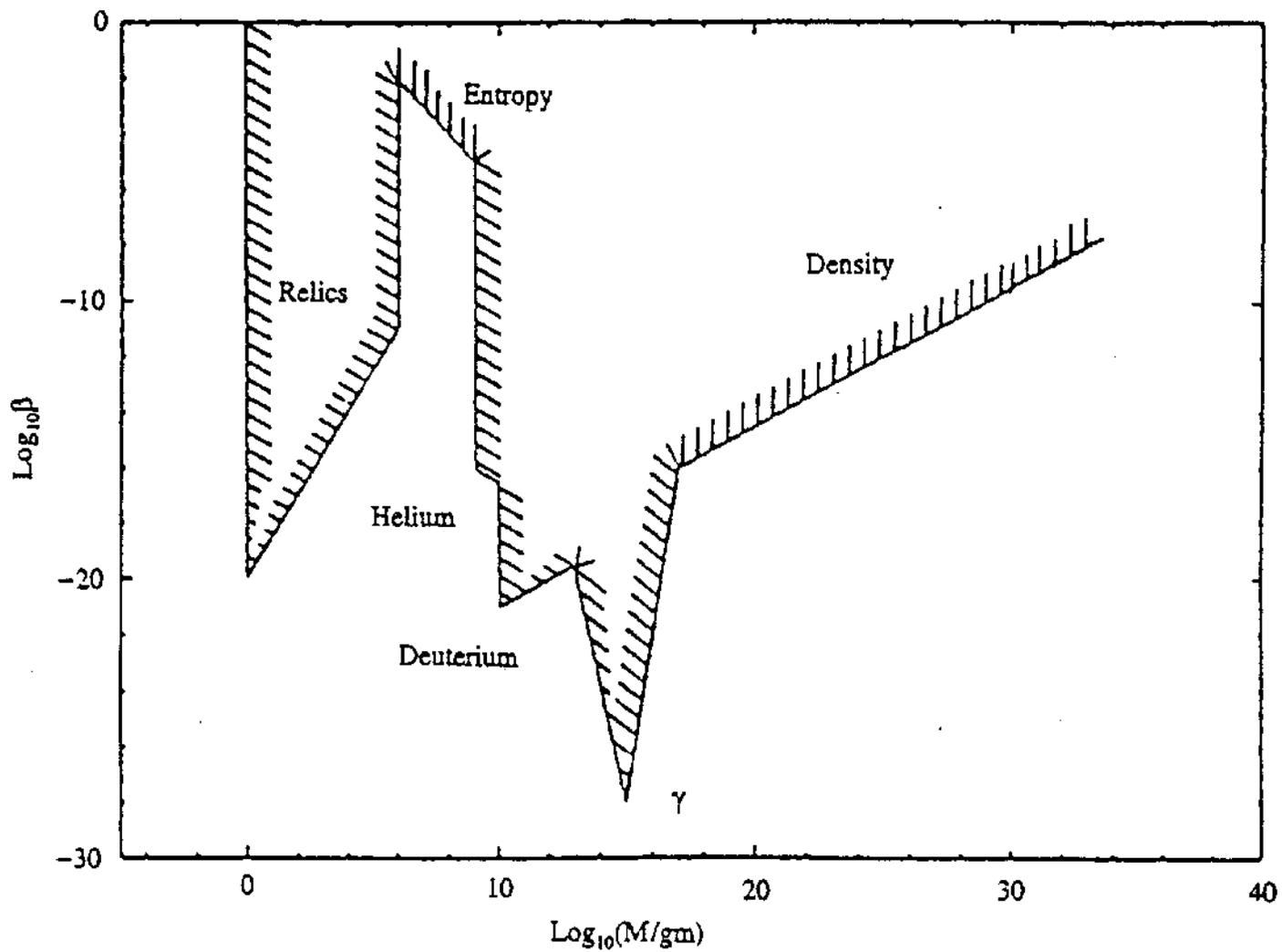
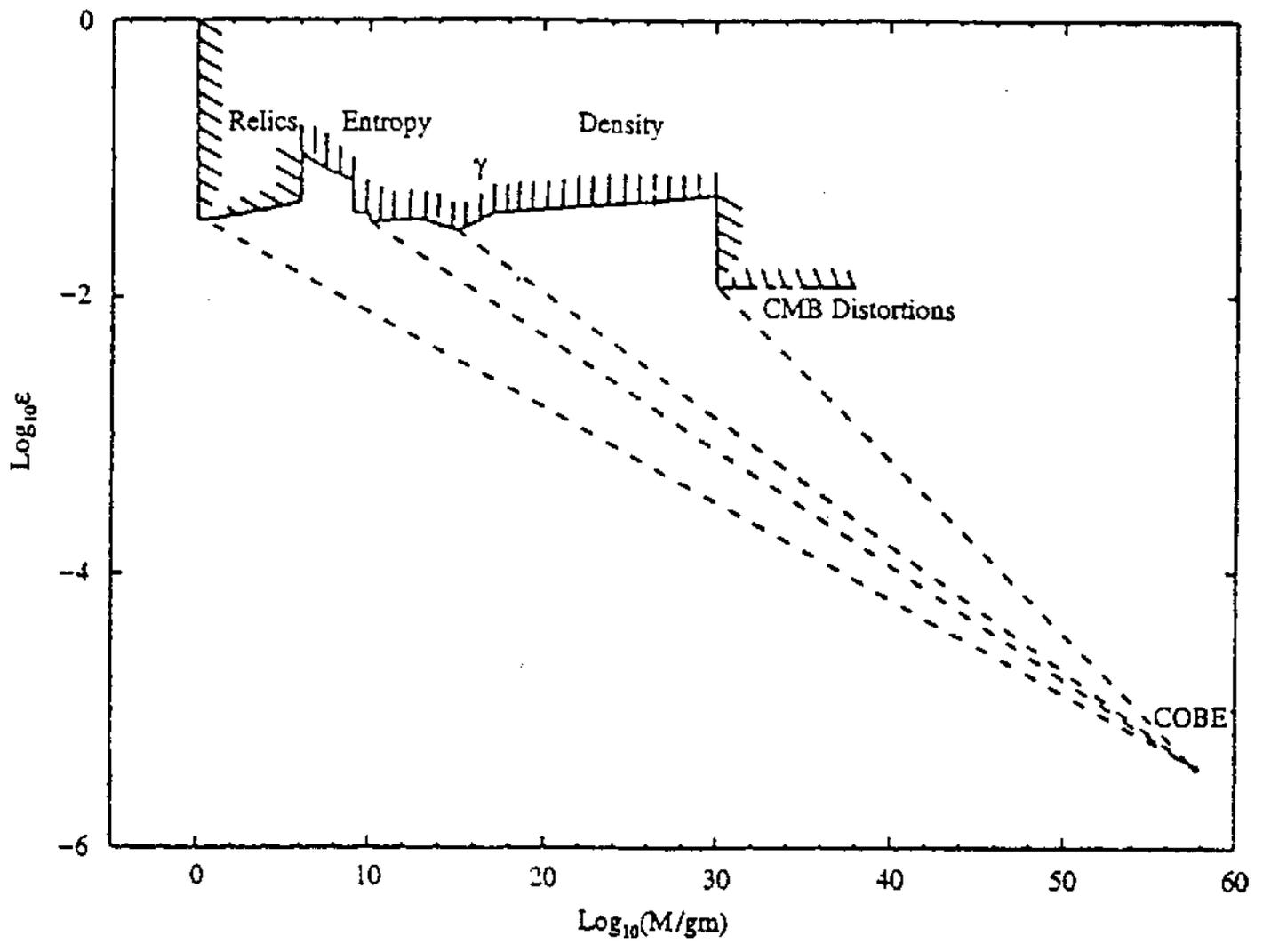
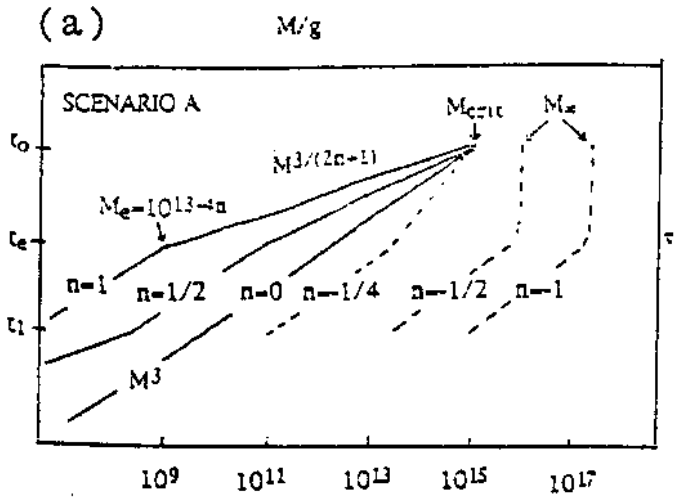
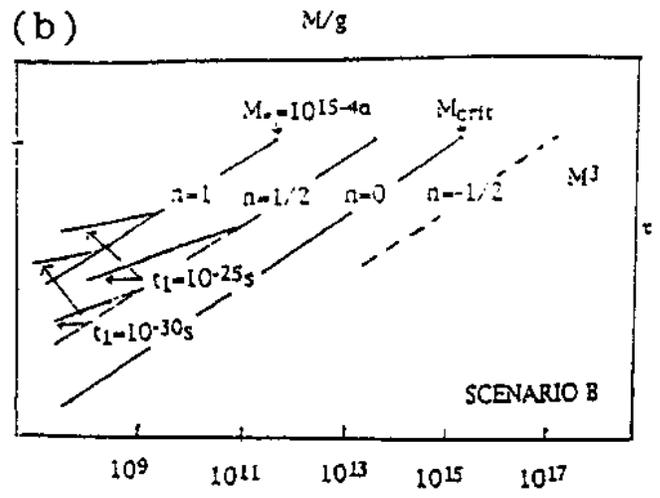
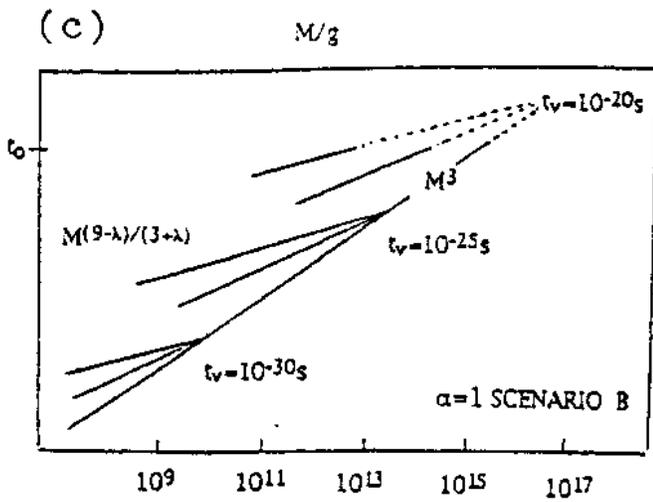
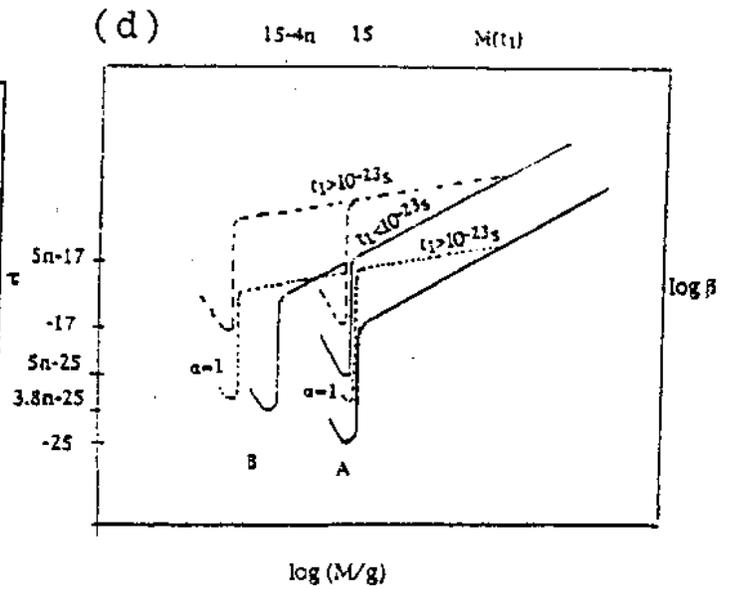
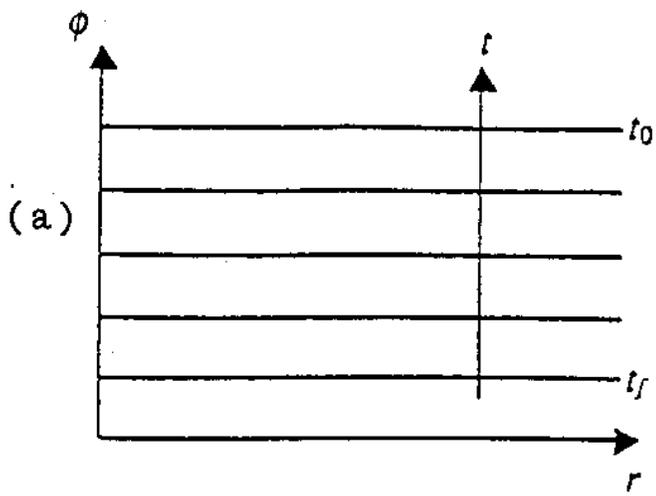
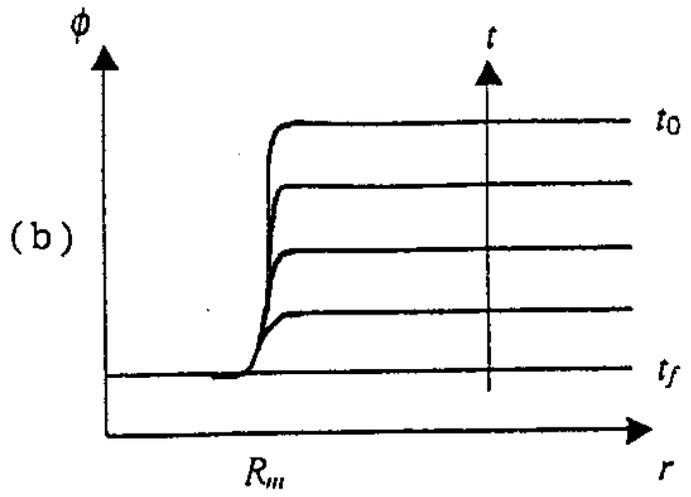
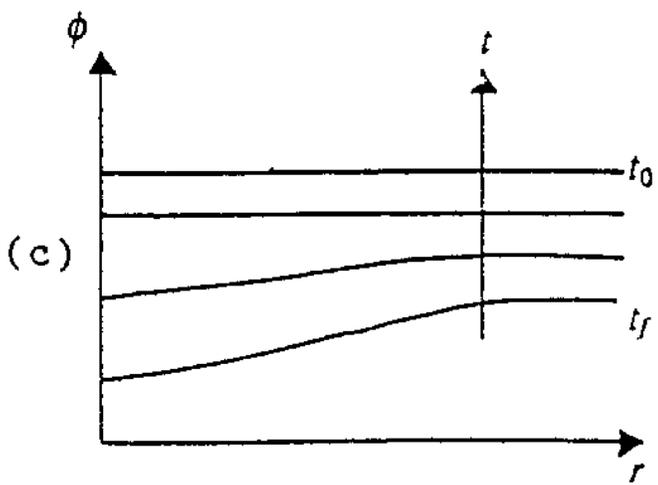
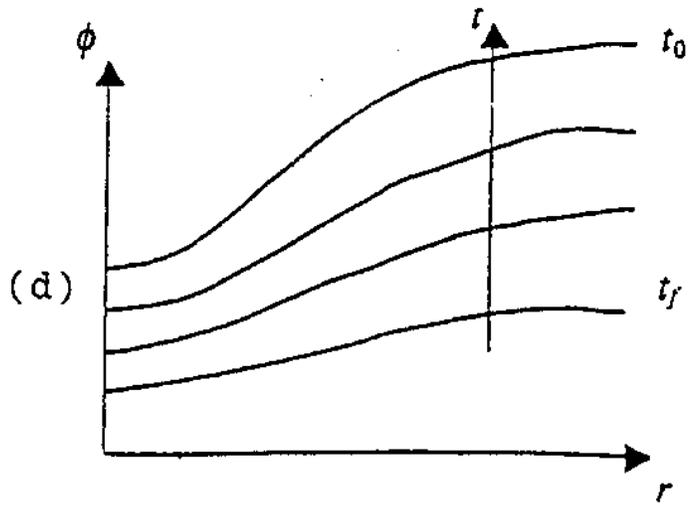
\end{document}